\documentclass[twocolumn,showpacs,preprintnumbers,amsmath,amssymb,floatfix]{revtex4}
\usepackage{graphicx}% Include figure files
\usepackage{bm}% bold math
\usepackage{psfrag} % psfrag for figures

\begin{document}

%\preprint{APS/123-QED}

\title{A microscopic approach to spin dynamics: about the meaning of spin relaxation times}

\author{C. Lechner}
%\email{christian.lechner@physik.uni-regensburg.de}
%\homepage{http://www.physik.uni-regensburg.de/forschung/roessler/}
\author{U. R\"ossler}
\affiliation{Institut f\"ur Theoretische Physik, Universit\"at Regensburg,
             D-93040 Regensburg}

\date{\today}

\begin{abstract}
We present an approach to spin dynamics by extending the optical Bloch equations for the driven two-level system to derive microscopic expressions for the transverse and longitudinal spin relaxation times. This is done for the 6-level system of electron and hole subband states in a semiconductor or a semiconductor quantum structure to account for the degrees-of-freedom of the carrier spin and the polarization of the exciting light and includes the scattering between carriers and lattice vibrations on a microscopic level. For the subsystem of the spin-split electron subbands we treat the electron-phonon interaction in second order and derive a set of equations of motion for the $2 \times 2$ spin-density matrix which describes the electron spin dynamics and contains microscopic expressions for the longitudinal ($T_1$) and the transverse ($T_2$) spin relaxation times. Their meaning will be discussed in relation to experimental investigations of these quantities. 
\end{abstract}

%\pacs{67.57.Lm,72.25.Rb,42.50.Md}

\maketitle

%%%%%%%%%%%%%%%%%%%%%%%%%%%%%%%%%%%%%%%%%%%%%%%%%%%%%%%%%%%%%%%%%%%%%%%%%%%%%%%%%%%%%%%%%%%%%%%%%%%%%%%%%%%
%begin of main text

\section{\label{introduction}Introduction}

The Bloch equations, originally formulated as equations of motion (EOM) for magnetic moments \cite{Bloch:1946} have turned out to apply in general to the dynamics of quantum mechanical two-level systems \cite{Cohenbook1:1977}. One prominent example are the \emph{optical Bloch equations} (OBE) in atomic or semiconductor physics with the components of the Bloch vector composed of the entries of the density matrix for a driven two-level system under excitation by a scalar light field (see e.g. Ref.~\onlinecite{Schaefer:2002}). Usually carrier scattering is accounted for by adding phenomenological damping terms connected with a longitudinal ($T_1$) and a transverse ($T_2$) relaxation time. In the context of OBE, $T_1$ characterizes the decay time of the population inversion or the relaxation into an equilibrium distribution, while $T_2$ is the timescale on which the coherence between exciting light and optical polarization gets lost. A further evolution of the OBE are the \emph{semiconductor Bloch equations} (SBE), which were formulated to describe optical phenomena in semiconductors under intense excitation by including many-particle terms due to Coulomb interaction between the carriers \cite{Lindberg:1988}. These equations yield a microscopic formulation of $T_1$ and $T_2$ caused by carrier-carrier \cite{Haug:1993} or carrier-phonon scattering \cite{Kuhn:1992,Schilp:1994}. In spite of their successful application to \emph{carrier dynamics}, the OBE and SBE, in their original form, are not capable to contribute to the current topic of \emph{spin dynamics} in semiconductors. Recently, this shortcoming was partially overcome by extending the SBE with respect to the spin degree-of-freedom of the carriers (including spin-orbit coupling) and the polarization degree-of-freedom of the exciting light \cite{Roessler:2002}, necessary to create a non-equilibrium spin distribution by optical orientation \cite{Meier:1984}. These extended SBE (derived by applying the Hartree-Fock truncation) are restricted, however, to the \emph{coherent regime} and hence fall short of describing scattering as origin of \emph{spin relaxation} and \emph{spin decoherence}, which are key issues of spintronics and quantum computation \cite{Awschalombook:2002}. On the other hand, the structure of these equations resembles those used in the phenomenological approach of spin dynamics \cite{Dyakonov:1971,Meier:1984, Averkiev:2002}, thus indicating the possibility of arriving at a microscopic approach to spin relaxation in the language of Bloch equations.

It is the aim of this paper to provide a microscopic formulation of spin dynamics in semiconductors and semiconductor heterostructures. We do this by starting from the extended OBE for the 6-level system of electron and hole subband states, containing the spin of the carriers and the polarization of the exciting light (this corresponds to taking into account only the single-particle contributions to the SBE of Ref.~\onlinecite{Roessler:2002}) and include the electron-phonon interaction as a possible scattering mechanism. 
For the subsystem of the conduction band states (spin-split by spin-orbit coupling) we formulate the full dynamics as a set of EOM for the $2\times2$ spin-density matrix and the phonon assisted density matrices. By using a correlation-based truncation scheme in second order Born approximation, we derive the scattering rates (in the Boltzmann limit and beyond) to arrive at equations describing the electron spin dynamics including relaxation (due to electron-phonon interaction) on a strictly microscopic level, while existing theories (see e.g. Ref.~\onlinecite{Averkiev:2002}) are a mixture of microscopics and phenomenology.
We want to stress also that, regarding the creation of a non-equilibrium spin population, our theoretical concept differs from some experimental situations: in our OBE a non-equilibrium spin polarization is due to \emph{optical orientation}, while in spintronic devices it is usually created by \emph{spin injection} \cite{Zutic:2004}. However, this difference will not become relevant in the context of this paper concentrating on the spin relaxation due to carrier-phonon interaction. 

This paper is organized as follows: In Sec.~\ref{untruncated} we introduce the system Hamiltonian, formulate the full dynamics of the system without truncation and derive the EOM for the electron subsystem. In Secs.~\ref{firstorder} and \ref{corrections} we present the correlation-based truncation scheme used to achieve a closed set of equations for the entries in the $2\times2$ density matrix related to the spin-split conduction band states. It represents an extension of the coherent OBE for the spin-density matrix by contributions due to electron-phonon scattering. In Sec.~\ref{unitarytrafo} we relate the dynamics of the density matrix with those of experimental observables and discuss the meaning of the corresponding spin relaxation times. Finally, we draw the conclusions of our results and give an outlook.             

%%%%%%%%%%%%%%%%%%%%%%%%%%%%%%%%%%%%%%%%%%%%%%%%%%%%%%%%%%%%%%%%%%%%%%%%%%%%%%%%%%%%%%%%%%%%%%%%%%%%%%%%%%%

\section{\label{untruncated}Spin-dependent OBE including carrier-phonon interaction}

\begin{figure}
\begin{psfrags}
\psfrag{k}{$\mathbf{k}$}
\psfrag{E}{$E$}
\psfrag{+k}{$+\mathbf{k}_0$}
\psfrag{-k}{$-\mathbf{k}_0$}
\psfrag{mc}{\tiny{$+m_c$}}
\psfrag{-mc}{\tiny{$-m_c$}}
\includegraphics[scale=0.2]{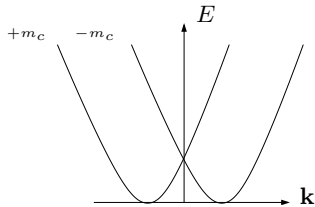}
\caption{\label{fig:Split} Sketch of subband splitting due to spin-orbit coupling.}
\end{psfrags}
\end{figure}

\begin{figure}
\begin{psfrags}
\psfrag{0}{$0$}
\psfrag{kx}{$k_x$}
\psfrag{ky}{$k_y$}
\includegraphics[scale=0.3]{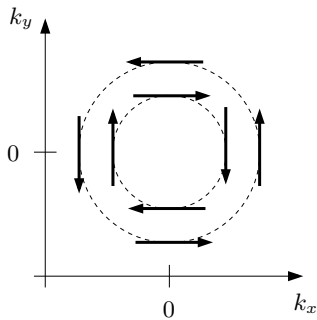}
\caption{\label{fig:Rashba} Sketch of constant energy contours of an electron subband (dashed lines), spin-split due to Rashba spin-orbit interaction. The dependence of the pseudospin orientation on the wave vector is visualized by the arrows.}
\end{psfrags}
\end{figure}

The Hamiltonian of the system is formulated in second quantization using the notation of Ref.~\onlinecite{Roessler:2002}. We restrict our discussion to the case of  a quantum well structure (QW), but the equations can be formulated in the same way for a bulk semiconductor. We consider a six-level system consisting of states from the spin-split lowest electron subband (with angular  momentum or pseudospin indices $m_c = \pm \frac{1}{2}$) and the corresponding heavy ($m_v=\pm\frac{3}{2}$) and light hole ($m_v = \pm\frac{1}{2}$) subband at wave vector $\mathbf{k}$ under excitation by a light field of arbitrary polarization and due to carrier-phonon interaction
\begin{equation}
\mathcal{H} = \mathcal{H}_{0} + \mathcal{H}_{light} + \mathcal{H}_{phonon}\; .
\label{sysham}
\end{equation}
In the following, we address the individual contributions to $\mathcal{H}$. Following Ref.~\onlinecite{Roessler:2002}, we adopt the diagonal form of
\begin{widetext}
\begin{equation}
\mathcal{H}_{0} = \sum_{\mathbf{k}'\, m_c'} \epsilon_{m_c'}(\mathbf{k}')\, c^{\dagger}_{m_c'}(\mathbf{k}')\, c_{m_c'}(\mathbf{k}') + \sum_{\mathbf{k}'\, m_v'} \epsilon_{m_v'}(\mathbf{k}')\, v_{m_v'}(\mathbf{k}')\, v^{\dagger}_{m_v'}(\mathbf{k}')\; ,
\label{hamdia}
\end{equation}
\end{widetext}
written with annihilation operators for electrons (holes) $c_{m_c'}(\mathbf{k}')$ $\big(v_{m_v'}(\mathbf{k}')\big)$ and corresponding creation operators. The time dependence of these operators is understood. The single-particle energies are denoted as $\epsilon_{m_c'}(\mathbf{k}')$ $\big(\epsilon_{m_v'}(\mathbf{k}')\big)$ for electrons (holes). Although the structure of these single-particle contributions is formally equivalent to a multisubband approach (see e.g. Ref. \onlinecite{Rossi:2002}) the physical content differs: In the multiband case the eigenenergies describe different bands separated by an energy gap (e.g. first and second electron subband). In contrast, we deal with subband states whose spin degeneracy is lifted due to $\mathbf{k}$-dependent spin-orbit coupling \cite{Roessler:1984, Lu:1998, Wissinger:1998} caused by \emph{bulk inversion asymmetry} (BIA or Dresselhaus term)\cite{Dresselhaus:1955} and/or \emph{structure inversion asymmetry} (SIA or Rashba term)\cite{Rashba:1960,Bychkov:1984} (see Fig.~\ref{fig:Split}), which are associated with spin precession. 
The diagonal form of $\mathcal{H}_{0}$ means that the angular momentum or \emph{pseudospin} is defined with respect to the direction of the wave vector $\mathbf{k}\,$. This particular dependence of the pseudospin orientation on the direction of $\mathbf{k}$ is visualized for the Rashba spin-orbit interaction in Fig.~\ref{fig:Rashba}.     
In dipole approximation, the interaction of the light field with the electrons and holes reads
\begin{widetext}
\begin{equation}
\mathcal{H}_{light} = - \sum_{m_c'\, m_v' \atop \mathbf{k}'} \left\{ \mathbf{E}(t)\cdot \mathbf{d}^{cv}_{m_c'\, m_v'}(\mathbf{k}')\,  
c^{\dagger}_{m_c'}(\mathbf{k}')\, v^{\dagger}_{m_v'}(\mathbf{k}') + \text{h.c.} \right\}\;,\\ 
\end{equation}
\end{widetext}
where $\mathbf{E}(t)$ is the electric field vector of the exciting light and $\mathbf{d}^{cv}_{m_c'\, m_v'}(\mathbf{k}')$ is the dipole matrix element between the two subband states with pseudospin index $m_c'$ and $m_v'\,$. The latter includes the optical selection rules (for details see Ref.~\onlinecite{Roessler:2002}). The vector notation is essential to account for the polarization degree-of-freedom which allows to create a non-equilibrium pseudospin distribution due to optical orientation \cite{Meier:1984}. 

\begin{figure}
\begin{psfrags}
\psfrag{0}{$0$}
\psfrag{kx}{$k_x$}
\psfrag{ky}{$k_y$}
\includegraphics[scale=0.3]{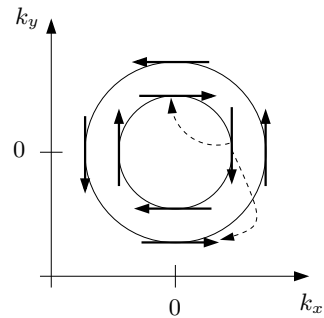}
\caption{\label{fig:Scatt} Possible scattering processes on the energy contours (thin full lines) connected with a change of the pseudospin.}
\end{psfrags}
\end{figure}

The Hamiltonian describing the phonons and the carrier-phonon interaction is given by
\begin{widetext}
\begin{eqnarray}
\nonumber
\mathcal{H}_{phonon} &=& \sum_{\mathbf{q}}\hbar\, \omega(\mathbf{q})\,b^{\dagger}(\mathbf{q})\, b(\mathbf{q})\\ \nonumber 
& & + \sum_{\mathbf{k}'\, \mathbf{q}}\left\{\sum_{m_c'\, m_c''} \left(g_{m_c'\, m_c''}^{e}(\mathbf{q})\, c^{\dagger}_{m_c''}(\mathbf{k}'+\mathbf{q})\, b(\mathbf{q})\, c_{m_c'}(\mathbf{k}') + \text{h.c.} \right)\right. \\
& &\mbox{} \hspace{0.9cm} + \left.\sum_{m_v'\, m_v''} \left(g_{m_v'\, m_v''}^{h}(\mathbf{q})\, v^{\dagger}_{m_v''}(\mathbf{k}'+\mathbf{q})\, b(\mathbf{q})\, v_{m_v'}(\mathbf{k}') + \text{h.c.} \right)\right\}\; ,
\label{phononham}
\end{eqnarray}
\end{widetext}
with the annihilation (creation) operator of a phonon $b^{(\dagger)}(\mathbf{q})\,$. The linear interaction of phonons with electrons (holes) is ruled by the matrix elements $g_{m_c'\, m_c''}^{e}(\mathbf{q})$ $\big(g_{m_v'\, m_v''}^{h}(\mathbf{q})\big)\,$, which due to our choice of the energy eigenstates as basis is labeled by pseudospin indices. This expresses the fact, that a change of the wave vector due to a scattering event is in general accompanied by a change of the pseudospin state as visualized in Fig. \ref{fig:Scatt}. We note that without spin-orbit coupling the eigenstates are simple products of orbital and spin (-up or -down) eigenstates. Consequently, the matrix elements of the (spin-conserving) electron-phonon interaction reduce to $g_{m_c'\,m_c''}= g\, \delta_{m_c'\,m_c''}\,$. The actual dependence of the interaction matrix element on $\mathbf{q}$ is determined by the interaction mechanism, which has to be specified for a quantitative analysis \cite{Kuhn:1992,Schilp:1994}. We note in passing, that the same system has been studied recently \cite{Weng:2003} even including besides electron-phonon interaction also impurity and carrier-carrier scattering, but not using the eigenstates of $\mathcal{H}_0\,$. Instead the spin-orbit coupling is treated there explicitly as an effective $\mathbf{k}$-dependent magnetic field giving rise to an inhomogeneous broadening. In Sec. \ref{unitarytrafo} we shall discuss this situation, which is related to the present choice of basis by a $\mathbf{k}$-dependent unitary transformation. 

In order to achieve expressions for the carrier-phonon scattering, we have to evaluate the EOM of the density matrix. In the basis of energy eigenfunctions of $\mathcal{H}_0$ the density matrix for the assumed 6-level model
\begin{equation}
\bm{\varrho}(\mathbf{k}) = 
\left(\begin{array} {c c}
\bm{\varrho}^{(m_c\, \bar{m}_c)}(\mathbf{k}) & \bm{\varrho}^{(m_c\, m_v)}(\mathbf{k}) \\
\bm{\varrho}^{(m_v\, m_c)}(\mathbf{k}) & \bm{\varrho}^{(m_v\, \bar{m}_v)}(\mathbf{k})   
\end{array} \right)
\label{densmatgen}  
\end{equation}
falls into different blocks, where $\bm{\varrho}^{(m_c\, \bar{m}_c)}(\mathbf{k})$ is the $2\times2$ matrix for the lowest conduction band and $\bm{\varrho}^{(m_v\, \bar{m}_v)}(\mathbf{k})$ represents the $4\times4$ matrix for the hole states. The off-diagonal blocks $\bm{\varrho}^{(m_c\, m_v)}(\mathbf{k})$ and $\bm{\varrho}^{(m_v\, m_c)}(\mathbf{k})$ include the coupling between valence and conduction band states by the exciting light field \cite{Roessler:2002}. They describe the interband or \emph{optical} coherence between the hole and electron states coupled by the light field and their time derivative defines the EOM of the interband polarization $ P_{m_c\, m_v}(\mathbf{k}) = \langle c^{\dagger}_{m_c}(\mathbf{k})\, v^{\dagger}_{m_v}(\mathbf{k}) \rangle\,$. Without electron-phonon interaction the EOM of all entries of the $6 \times 6$ density matrix would form a closed set of equations representing the \emph{coherent} spin-dependent OBE for the system. A detailed theoretical study of the optical coherence and polarization dynamics, yet without addressing the spin/pseudospin, can be found in Ref.~\onlinecite{Rossi:2002}. 

The pseudospin dynamics, in particular the relaxation and decoherence, is contained in the time evolution of the diagonal blocks, which shall be exemplified here for the electron system. The same steps of calculation would lead to the corresponding equations for the hole system, which however are more complicated due to the additional orbital degrees-of-freedom. The $2\times2$ pseudospin-density matrix for the electrons is
\begin{equation}
\bm{\varrho}^{(m_c\, \bar{m}_c)}(\mathbf{k}) = 
\left(\begin{array}{c c}
\varrho_{m_c\, m_c}(\mathbf{k}) & \varrho_{m_c\, -m_c}(\mathbf{k}) \\
\varrho_{-m_c\, m_c}(\mathbf{k}) & \varrho_{-m_c\, -m_c}(\mathbf{k}) 
\end{array} \right)\; .
\label{denscond}
\end{equation}
The single entries are expectation values of products of a creation and an annihilation operator $\varrho_{m_c\, \bar{m}_c}(\mathbf{k})= \langle c^{\dagger}_{m_c}(\mathbf{k})\, c_{\bar{m}_c}(\mathbf{k}) \rangle\,$. We evaluate the commutators of the system Hamiltonian, Eq. \eqref{sysham}, with $c^{\dagger}_{m_c}(\mathbf{k})\,c_{\bar{m}_c}(\mathbf{k})$ and take the thermal expectation value to find their EOM
\begin{widetext}
\begin{eqnarray}
\nonumber
i \hbar \partial_t\, \varrho_{m_c\, \bar{m}_c}(\mathbf{k}) &=& 
\left(\epsilon_{m_c}(\mathbf{k}) - \epsilon_{\bar{m}_c}(\mathbf{k})\right)\, \varrho_{m_c\, \bar{m}_c}(\mathbf{k}) \\ \nonumber
&& + \sum_{m_v} \left\{ \mathbf{E}(t)\cdot \mathbf{d}^{cv}_{\bar{m}_c\, m_v}(\mathbf{k})\, P_{m_c\, m_v}(\mathbf{k}) -  \mathbf{E}^{*}(t)\cdot \mathbf{d}^{cv\,*}_{m_c\, m_v}(\mathbf{k})\, P^{\dagger}_{\bar{m}_c\, m_v}(\mathbf{k})\right\}\\ \nonumber
&&+ \sum_{\mathbf{q}\, m_{c}'}\bigg\{g_{m_c'\, m_c}^{e}(\mathbf{q}) \langle c_{m_c'}^{\dagger}(\mathbf{k} + \mathbf{q})\, b(\mathbf{q})\, c_{\bar{m}_c}(\mathbf{k}) \rangle  - g_{\bar{m}_c\, m_c'}^{e}(\mathbf{q}) \langle c_{m_c}^{\dagger}(\mathbf{k})\, b(\mathbf{q})\, c_{m_c'}(\mathbf{k} - \mathbf{q}) \rangle \\ \nonumber
&&\mbox{} \hspace{0.8cm} + g_{m_c\, m_c'}^{e\, *}(\mathbf{q}) \langle c_{m_c'}^{\dagger}(\mathbf{k}- \mathbf{q})\, b^{\dagger}(\mathbf{q})\, c_{\bar{m}_c}(\mathbf{k}) \rangle  - g_{m_c'\, \bar{m}_c}^{e\, *}(\mathbf{q}) \langle c_{m_c}^{\dagger}(\mathbf{k})\, b^{\dagger}(\mathbf{q})\, c_{m_c'}(\mathbf{k} + \mathbf{q}) \rangle \bigg\}\; . \\ 
\label{eomrhooff}
\end{eqnarray}
\end{widetext}
The first two lines are the single-particle contributions of the SBE in Ref. \cite{Roessler:2002}: they describe the dynamics caused by the spin-split energy levels and by the excitation of the electrons of either pseudospin from the valence subbands depending on the polarization of the driving light field. The three-operator terms specify the scattering of an electron (in one of the spin-split subbands) from one $\mathbf{k}$ to another one (in the same or the other spin-split subband) thereby absorbing or emitting a phonon, as visualized in Fig.~\ref{fig:Scatt}. The three-operator terms (or their thermal expectation values) establish the phonon-assisted density matrix \cite{Kuhn:1992}, whose entries obey EOMs of which we present as an example the one for $s_{m_c'\, \bar{m}_c}(\mathbf{k}+\mathbf{q},\, \mathbf{q}) = \langle c^{\dagger}_{m_c'}(\mathbf{k}+\mathbf{q})\, b(\mathbf{q})\, c(\mathbf{k}_{\bar{m}_c}) \rangle$
\begin{widetext}
\begin{eqnarray}
\nonumber
i \hbar \partial_t\,  s_{m_c'\, \bar{m}_c}(\mathbf{k}+\mathbf{q},\, \mathbf{q})
&=& \left(\epsilon_{m_c'}(\mathbf{k}+\mathbf{q}) - \epsilon_{\bar{m}_c}(\mathbf{k}) - \hbar \omega(\mathbf{q}) \right)\, s_{m_c'\, \bar{m}_c}(\mathbf{k}+\mathbf{q},\, \mathbf{q}) \\ \nonumber
&& + \sum_{\mathbf{k}'\, \mathbf{q}'}\sum_{\tilde{m}_c \, \tilde{m}_c' \atop \tilde{m}_v \, \tilde{m}_v'} \left\{g^{e}_{\tilde{m}_c'\, m_c'}(\mathbf{q}')\langle c^{\dagger}_{\tilde{m}_c'}(\mathbf{k}+ \mathbf{q} + \mathbf{q}')\, b(\mathbf{q}')\, b(\mathbf{q})\, c_{\bar{m}_c}(\mathbf{k}) \rangle \right. \\ \nonumber
&&\mbox{} \hspace{1.8cm}- g^{e}_{\bar{m}_c\, \tilde{m}_c}(\mathbf{q}')\langle c^{\dagger}_{m_c'}(\mathbf{k}+ \mathbf{q})\, b(\mathbf{q}')\, b(\mathbf{q})\, c_{\tilde{m}_c}(\mathbf{k}- \mathbf{q}') \rangle \\ \nonumber
&&\mbox{} \hspace{1.8cm} + g^{e\,*}_{m_c'\, \tilde{m}_c}(\mathbf{q}')\langle c^{\dagger}_{\tilde{m}_c}(\mathbf{k} + \mathbf{q} - \mathbf{q}')\, b^{\dagger}(\mathbf{q}')\, b(\mathbf{q})\, c_{\bar{m}_c}(\mathbf{k}) \rangle \\ \nonumber
&&\mbox{} \hspace{1.8cm} - g^{e\,*}_{\tilde{m}_c'\, \bar{m}_c}(\mathbf{q}')\langle c^{\dagger}_{m_c'}(\mathbf{k}+\mathbf{q})\, b(\mathbf{q})\, b^{\dagger}(\mathbf{q}')\, c_{\tilde{m}_c'}(\mathbf{k} + \mathbf{q}') \rangle \\ \nonumber
&&\mbox{} \hspace{1.8cm} + g^{e\,*}_{\tilde{m}_c'\, \tilde{m}_c}(\mathbf{q})\langle c^{\dagger}_{\tilde{m}_c}(\mathbf{k}')\, c^{\dagger}_{m_c'}(\mathbf{k}+\mathbf{q})\, c_{\tilde{m}_c'}(\mathbf{k}' + \mathbf{q})\, c_{\bar{m}_c}(\mathbf{k}) \rangle \\ \nonumber
&&\mbox{} \hspace{1.8cm} \left. - g^{h}_{\tilde{m}_v'\, \tilde{m}_v}(\mathbf{q})\langle v^{\dagger}_{\tilde{m}_v}(\mathbf{k}')\, c^{\dagger}_{m_c'}(\mathbf{k}+\mathbf{q})\, v_{\tilde{v}'}(\mathbf{k}' + \mathbf{q}')\, c_{\bar{m}_c}(\mathbf{k}) \rangle \right\}\; .\\
\label{eomphonass1}
\end{eqnarray}
\end{widetext}
As can be seen from Eqs. \eqref{eomrhooff} and \eqref{eomphonass1}, we run into a hierarchy problem with EOMs containing terms with an increasing number of operators, which is typical for systems with interactions. This hierarchy problem can be overcome by a proper truncation. The standard procedure is to neglect the existence of coherent phonons corresponding to the expectation value of a single bosonic operator (first order factorization) and to take into account only the expectation values which lead to a phonon occupation number \cite{Kuhn:1992,Schilp:1994}.

%%%%%%%%%%%%%%%%%%%%%%%%%%%%%%%%%%%%%%%%%%%%%%%%%%%%%%%%%%%%%%%%%%%%%%%%%%%%%%%%%%%%%%%%%%%%%%%%%%%%%%%%%%%
\section{\label{firstorder}The Boltzmann limit}

The goal of the truncation is to gradually filter out the scattering terms up to a certain order in the interaction relevant for the investigated dynamics. To express the scattering contributions in the \emph{Boltzmann limit} caused by electron-phonon interaction, we formulate the following rules for the truncation:
\begin{enumerate}
\item After factorization of the four-operator terms only expressions containing a macroscopic expectation value are taken into account.
\item Scattering terms contributing in the Boltzmann limit are those proportional to the squared absolute value of the interaction matrix element in Eq.~\eqref{eomrhooff}. This means that we neglect the so-called ``polarization scattering'' due to inter- and intraband processes \cite{Kutznetsov:1991} for which we refer to the next section.
\end{enumerate}

Applying these rules modifies Eq.~\eqref{eomphonass1} and leads to an equation with the following characteristic structure
\begin{equation}
\partial_t\, x(t) = - i\, \omega\, x(t) + y(t)\; ,
\label{integrationa}
\end{equation}
where $x(t)$ stands for the three-operator term and $y(t)$ corresponds to products between phonon occupation functions $\beta(\mathbf{q})=\langle b^{\dagger}(\mathbf{q})\, b(\mathbf{q}) \rangle$ and entries of the electron density matrix $\varrho_{m_c\, \bar{m}_c}(\mathbf{k}')$.
As presented in Ref.~\onlinecite{Kuhn:1992}, equations of this type can formally be integrated to yield
\begin{equation}
x(t) = x(t_0) e^{-i \omega(t-t_0)} + \int_{0}^{t-t_0} e^{-i\omega\, t'}\, y(t-t')\, dt'\;.
\label{integrationb}
\end{equation}
Inserting this result into the EOM of $\varrho_{m_c\, \bar{m}_c}(\mathbf{k})$ leads to a non-Markovian integro-differential equation, which can be solved analytically by applying the Markov and adiabatic limit \cite{Zimmermann:1990,Kutznetsov:1991}. This corresponds to use instead of Eq. \eqref{integrationb} the following expression
\begin{equation}
x(t) = \left(-i \frac{\mathcal{P}}{\omega} + \pi \delta(\omega)\right)\, y(t)\; ,
\label{integrationc}
\end{equation}
where $\mathcal{P}$ denotes the principal value. With this approach we solve the EOM for the different entries of the phonon-assisted density matrices appearing in Eq.~\eqref{eomrhooff}. 

As we are interested in the relaxation due to electron-phonon interaction we write down only the scattering contributions to the EOMs of the entries of the pseudospin-density matrix, Eq. \eqref{denscond}, which correspond to damping terms. We obtain for the diagonal entries
\begin{widetext}
\begin{equation}
\partial_t\, \varrho_{m_c\, m_c}(\mathbf{k})\pmb{\vert}_{scatt1}= -\Gamma^{out}_{m_c\,m_c}(\mathbf{k})\, \varrho_{m_c\,m_c}(\mathbf{k}) + \Gamma^{in}_{m_c\,m_c}(\mathbf{k})\, \left(1-\varrho_{m_c\,m_c}(\mathbf{k})\right)\; ,
\label{finalrhodia}
\end{equation}
\end{widetext}
with the characteristic rates for scattering ``out'' of or ``in'' to the state with pseudospin $m_c$ and wavevector $\mathbf{k}$. The explicit form of $\Gamma^{out}_{m_c\,m_c}(\mathbf{k})$ is given by
\begin{widetext}
\begin{eqnarray}
\nonumber
\Gamma^{out}_{m_c\,m_c}(\mathbf{k}) &=& \frac{\pi}{\hbar}\, \sum_{\mathbf{q}, m_c'} |g_{m_c'\,m_c}(\mathbf{q})|^2\,\Big\{ \delta\left(\epsilon_{m_c'}(\mathbf{k}+ \mathbf{q}) - \epsilon_{m_c}(\mathbf{k}) - \hbar \omega(\mathbf{q})\right)\, \left(1 - \varrho_{m_c'\, m_c'}(\mathbf{k}+ \mathbf{q})\right)\, \beta(\mathbf{q})  \\ \nonumber
&&\mbox{} \hspace{1.6cm} + \delta\left(\epsilon_{m_c'}(\mathbf{k} - \mathbf{q}) - \epsilon_{m_c}(\mathbf{k}) + \hbar \omega(\mathbf{q}) \right)\, \left(1 - \varrho_{m_c'\, m_c'}(\mathbf{k}- \mathbf{q})\right)\,\left(1 + \beta(\mathbf{q})\right)\Big\}\;.\\ 
\end{eqnarray}
\end{widetext}
It has the characteristic form of expressions obtained from Fermi's Golden Rule: all terms are proportional to the absolute squared value of the interaction matrix elements and to the $\delta$-function to warrant energy conservation in the scattering process. $\Gamma^{in}_{m_c\,m_c}(\mathbf{k})$ has the same form but with changed phonon and electron occupation factors.

For the off-diagonal entries we may write the scattering contributions as
\begin{equation}
\partial_t\, \varrho_{m_c\, -m_c}(\mathbf{k})\pmb{\vert}_{scatt1} = - \Gamma^{e-p}_{m_c\, -m_c}(\mathbf{k})\, \varrho_{m_c\, -m_c}(\mathbf{k})\; , 
\label{finalrhooff}
\end{equation}
with 
\begin{widetext}
\begin{eqnarray}
\nonumber
\Gamma^{e-p}_{m_c\, -m_c}(\mathbf{k}) &=& \frac{\pi}{\hbar}\sum_{\mathbf{q}\, m_c'}\Big\{|g^{e}_{m_c'\, m_c}(\mathbf{q})|^2 \, \delta\left(\epsilon_{m_c'}(\mathbf{k}+\mathbf{q}) - \epsilon_{-m_c}(\mathbf{k})   - \hbar \omega(\mathbf{q})\right)\,\times \\ \nonumber
&& \left[\left(1 - \varrho_{m_c'\, m_c'}(\mathbf{k} + \mathbf{q})\right)\, \beta(\mathbf{q}) + \varrho_{m_c'\, m_c'}(\mathbf{k}+ \mathbf{q})\left(1 + \beta(\mathbf{q})\right) \right]\\ \nonumber
&&+|g^{e}_{m_c'\, m_c}(\mathbf{q})|^2 \, \delta\left(\epsilon_{m_c'}(\mathbf{k}-\mathbf{q})-\epsilon_{-m_c}(\mathbf{k}) + \hbar \omega(\mathbf{q})\right)\,\times \\ \nonumber
&& \left[\left(1 - \varrho_{m_c'\, m_c'}(\mathbf{k} - \mathbf{q})\right)\,\left( 1 + \beta(\mathbf{q})\right) + \varrho_{m_c'\, m_c'}(\mathbf{k}- \mathbf{q})\, \beta(\mathbf{q}) \right] \\ \nonumber
&&+|g^{e}_{m_c'\, -m_c}(\mathbf{q})|^2 \, \delta \left(\epsilon_{m_c'}(\mathbf{k} + \mathbf{q}) - \epsilon_{m_c}(\mathbf{k}) - \hbar \omega(\mathbf{q})\right)\,\times \\ \nonumber
&& \left[\left(1 - \varrho_{m_c'\, m_c'}(\mathbf{k} + \mathbf{q})\right)\,\beta(\mathbf{q}) + \varrho_{m_c'\, m_c'}(\mathbf{k} + \mathbf{q})\, \left(1+\beta(\mathbf{q})\right) \right] \\ \nonumber
&&+|g^{e}_{m_c'\, -m_c}(\mathbf{q})|^2 \, \delta \left(\epsilon_{m_c'}(\mathbf{k} - \mathbf{q}) - \epsilon_{m_c}(\mathbf{k}) + \hbar \omega(\mathbf{q})\right)\,\times \\ \nonumber
&& \left[\left(1 - \varrho_{m_c'\, m_c'}(\mathbf{k} - \mathbf{q})\right)\,\left( 1 + \beta(\mathbf{q})\right) + \varrho_{m_c'\, m_c'}(\mathbf{k}- \mathbf{q})\, \beta(\mathbf{q}) \right] \Big\}\; .\label{GammaBoltzmann} \\
\end{eqnarray}
\end{widetext}

Following the line of arguments in Ref. \onlinecite{Lindberg:1988} or \onlinecite{Haug:1993}, where the SBE have been derived for the two-level system of a conduction and a valence band state with carrier-carrier interaction, we may identify the damping rates in our pseudospin system with the inverse relaxation times
\begin{eqnarray}
\frac{1}{T_{1,\, \mathbf{k}}} &=& \sum_{m_c'} \left(\Gamma^{in}_{m_c'\, m_c'}(\mathbf{k}) + \Gamma^{out}_{m_c'\, m_c'}(\mathbf{k})\right)\\
\frac{1}{T_{2,\, \mathbf{k}}} &=& \Gamma^{e-p}_{m_c\,-m_c}(\mathbf{k})
\end{eqnarray}
of a Bloch vector, whose components are defined in the usual way by entries of the $2\times2$ pseudospin-density matrix. In spite of the similarities in the microscopic expressions of $T_{1,\, \mathbf{k}}$ and $T_{2,\, \mathbf{k}}\,$ for our system, we do not find the relation $2\, T_{1,\, \mathbf{k}}=T_{2,\, \mathbf{k}}$ as for the system studied in Ref. \onlinecite{Haug:1993}. We notice instead a one-to-one correspondence between the individual contributions to both rates except for the sign changes in the pseudospin index $m_c\,$, which -- given the small spin-splitting -- leads to a relation $T_{1,\, \mathbf{k}} \simeq T_{2,\, \mathbf{k}}\,$. In fact for a system with spin-degenerate electron states, $\epsilon_{m_c}(\mathbf{k})=\epsilon_{-m_c}(\mathbf{k})$ (as in systems with inversion symmetry), we find exactly $T_{1,\, \mathbf{k}} = T_{2,\, \mathbf{k}}\,$. This is in accordance with the argument used for two-level systems (see e.g. chapter 4 of Ref. \onlinecite{Awschalombook:2002}), that a significant difference in these times ($T_{1} \gg T_{2}\,$) arises, if due to separation of the two levels the decay of the population inversion requires energy dissipation. Making the \emph{gedanken} experiment by assuming that the spin-orbit interaction is completely  ``switched-off'', the pseudospin and the orbital degree-of-freedom decouple. Hence, all dependencies on pseudospin indices are redundant as only \emph{spin-conserving} scattering processes are possible. As a consequence no relaxation to a pseudospin-equilibrium is possible, because the scattering due to phonons is no longer capable of changing the pseudospin orientation. Nevertheless the scattering does not vanish but leads to a redistribution of the states in $\mathbf{k}$-space \emph{within} the separate pseudospin reservoirs.      
    
%  
%%%%%%%%%%%%%%%%%%%%%%%%%%%%%%%%%%%%%%%%%%%%%%%%%%%%%%%%%%%%%%%%%%%%%%%%%%%%%%%%%%%%%%%%%%%%%%%%%%%%%%%%%%%

\section{\label{corrections}Beyond the Boltzmann limit}

In order to include all scattering terms up to second order in the electron-phonon interaction we have to go \emph{beyond the Boltzmann limit} by taking into account also those contributions to the EOM of the phonon-assisted density matrices (taking Eq. \eqref{eomphonass1} as an example), which were omitted in the previous section. This is achieved by relaxing the second truncation rule used in Sec. \ref{firstorder} and leads to additional contributions only to the EOM of the off-diagonal entry of the $2\times2$ electron pseudospin-density matrix which can be cast into the form
\begin{equation}
\partial_t\, \varrho_{m_c\, -m_c}(\mathbf{k})\pmb{\vert}_{scatt2} = -\frac{1}{i \hbar}\,\sum_{\mathbf{q}\, m_c'} \bar{\Sigma}^{e-p}_{m_c'\, -m_c'}(\mathbf{q})\, \varrho_{m_c'\, -m_c'}(\mathbf{k} + \mathbf{q})\; . 
\label{eomrhooffboltz}
\end{equation}
In contrast to Eq. \eqref{finalrhooff} one has to sum here over the pseudospin index and the wave vector which enters differently in the self-energy $\bar{\Sigma}^{e-p}_{m_c'\, -m_c'}(\mathbf{q})$ and in $\varrho_{m_c'\, -m_c'}(\mathbf{k} + \mathbf{q})\,$. A corresponding scattering contribution was found in Ref. \cite{Haug:1993,Kuhn:1992} for the interband polarization, i.e. for the off-diagonal entry of the $2\times2$ density matrix considered there. In order to present the structure of the self-energy we extract all contributions containing (according to Eq. \eqref{integrationc}) a $\delta$-function by writing
\begin{widetext}
\begin{eqnarray}
\nonumber
\bar{\Gamma}^{e-p}_{m_c'\, -m_c'}(\mathbf{q}) &=& 
\frac{\pi}{\hbar}\Big\{
g^{e}_{-m_c\, -m_c'}(\mathbf{q})\, g^{e\, *}_{m_c\, m_c'}(\mathbf{q}) \, \delta\left(\epsilon_{-m_c'}(\mathbf{k}+\mathbf{q}) -\epsilon_{m_c}(\mathbf{k})+ \hbar \omega(\mathbf{q})\right)\,\times \\ \nonumber
&& \left[\left(1 - \varrho_{m_c\, m_c}(\mathbf{k})\right)\, \beta(\mathbf{q}) + \varrho_{m_c\, m_c}(\mathbf{k})\,\left( 1 + \beta(\mathbf{q})\right) \right] \\ \nonumber
&&+ g^{e}_{-m_c\, -m_c'}(\mathbf{q})\, g^{e\,*}_{m_c\, m_c'}(\mathbf{q}) \, \delta \left(\epsilon_{m_c'}(\mathbf{k} + \mathbf{q})-\epsilon_{-m_c}(\mathbf{k}) + \hbar \omega(\mathbf{q})\right)\,\times \\ \nonumber
&& \left[\left(1 - \varrho_{-m_c\, -m_c}(\mathbf{k})\right)\,\beta(\mathbf{q}) + \varrho_{-m_c\, -m_c}(\mathbf{k})\, \left( 1 + \beta(\mathbf{q})\right) \right] \\ \nonumber
&&+ g^{e}_{m_c'\, m_c}(\mathbf{q})\, g^{e\, *}_{-m_c'\, -m_c}(\mathbf{q}) \, \delta\left(\epsilon_{m_c'}(\mathbf{k}+\mathbf{q}) - \epsilon_{-m_c}(\mathbf{k}) - \hbar \omega(\mathbf{q})\right)\,\times \\ \nonumber
&& \left[(1- \varrho_{-m_c\, -m_c}(\mathbf{k}))\left(1 + \beta(\mathbf{q})\right) + \varrho_{-m_c\, -m_c}(\mathbf{k})\, \beta(\mathbf{q}) \right]\\ \nonumber
&&+ g^{e}_{m_c'\, m_c}(\mathbf{q})\, g^{e\, *}_{-m_c'\, -m_c}(\mathbf{q}) \, \delta \left(\epsilon_{-m_c'}(\mathbf{k} + \mathbf{q}) - \epsilon_{m_c}(\mathbf{k}) - \hbar \omega(\mathbf{q})\right)\,\times \\
&& \left[\left(1 - \varrho_{m_c\, m_c}(\mathbf{k})\right)\,\left(1+\beta(\mathbf{q})\right) + \varrho_{m_c\, m_c}(\mathbf{k})\, \beta(\mathbf{q}) \right]\Big\}\; . \label{GammaBeyond}  
\end{eqnarray}
\end{widetext}
We can identify the source terms composed of products of distribution functions and related to the different scattering processes. As before, the energy conservation is contained in the delta function (therefore, we have omitted the contribution with $g^{h}_{\tilde{m}_v'\, \tilde{m}_v}(\mathbf{q})$ as a factor, because the energy difference between conduction and valence band states is usually much larger than the phonon energy). In contrast to Eq. \eqref{GammaBoltzmann}, the terms are not proportional to the absolute squared values of the interaction matrix elements, which is typical for the contributions beyond the Boltzmann limit as can be seen by comparing with the corresponding result for the two-level system of Ref. \onlinecite{Kuhn:1992} which shows a similar structure. There the terms beyond the Boltzmann limit have been denoted as polarization scattering \cite{Kutznetsov:1991} with reference to the interband polarization, while here they mean the corresponding scattering processes in the dynamics of $\varrho_{m_c\, -m_c}(\mathbf{k})\,$.

Together with the results of the previous sections we may now write the full set of EOMs for the pseudospin-density matrix
\begin{widetext}
\begin{eqnarray}
\nonumber
\partial_t\, \varrho_{m_c\,m_c}(\mathbf{k}) &=& \frac{1}{i\, \hbar}\sum_{m_v} \left\{\mathbf{E}(t)\cdot \mathbf{d}^{cv}_{m_c\, m_v}(\mathbf{k})\, P_{m_c\, m_v}(\mathbf{k}) - \text{h.c.} \right\}
 \\ 
&& - \Gamma^{out}_{m_c\, m_c}(\mathbf{k})\, \varrho_{m_c\, m_c}(\mathbf{k})  +\Gamma^{in}_{m_c\,-m_c}(\mathbf{k})\, \left(1- \varrho_{m_c\, m_c}(\mathbf{k})\right)\label{DPfinaldia}\\ \nonumber
&& \\ \nonumber
\partial_t\, \varrho_{m_c\,-m_c}(\mathbf{k}) &=& \frac{1}{i\,\hbar}\, \left(\epsilon_{m_c}(\mathbf{k}) - \epsilon_{-m_c}(\mathbf{k})\right)\, \varrho_{m_c\, -m_c}(\mathbf{k}) \\ \nonumber
&&+ \frac{1}{i\, \hbar}\sum_{m_v} \left\{\mathbf{E}(t)\cdot \mathbf{d}^{cv}_{-m_c\, m_v}(\mathbf{k})\, P_{m_c\, m_v}(\mathbf{k}) -  \mathbf{E}^{*}(t)\cdot \mathbf{d}^{cv\,*}_{m_c\, m_v}(\mathbf{k})\, P^{\dagger}_{-m_c\, m_v}(\mathbf{k})\right\}
 \\ 
&& - \Gamma^{e-p}_{m_c\, -m_c}(\mathbf{k})\, \varrho_{m_c\, -m_c}(\mathbf{k})  +\,\sum_{\mathbf{q}\, m_c'} \bar{\Sigma}^{e-p}_{m_c'\, -m_c'}(\mathbf{q})\, \varrho_{m_c'\, -m_c'}(\mathbf{k} + \mathbf{q})\; .
\label{DPfinaloff}
\end{eqnarray}
\end{widetext}
It contains in a microscopic formulation the pseudospin dynamics in electron subbands due to spin-orbit coupling, spin-selective optical excitation and electron-phonon interaction (for carrier-carrier interaction see the remark at the end of this paper). By properly defining a Bloch vector as in Sec. \ref{firstorder} and looking at the damping terms in the corresponding Bloch equations we can again specify the longitudinal and transverse pseudospin relaxation times. As it turned out, only $T_{2,\, \mathbf{k}}$ is modified by additional terms (beyond the Boltzmann limit) discussed in this section, while $T_{1,\, \mathbf{k}}$ remains unchanged.

%%%%%%%%%%%%%%%%%%%%%%%%%%%%%%%%%%%%%%%%%%%%%%%%%%%%%%%%%%%%%%%%%%%%%%%%%%%%%%%%%%%%%%%%%%%%%%%%%%%%%%%%%%%

\section{\label{unitarytrafo}Changing the spin basis}

When describing experiments designed to measure the \emph{spin relaxation} time $\tau_{SR}$ and the \emph{spin decoherence} time $\tau_{SD}$ of a system (see e.g. Ref.~\onlinecite{Zutic:2004} and references therein), a basis is used with spin states oriented relative to a fixed direction, e.g. the growth direction of the QW structure. According to this choice, spins are spin-up ($\uparrow$) or spin-down ($\downarrow$) when aligned parallel or antiparallel to this direction, but in the presence of spin-orbit interaction spin is not a good quantum number. Consequently, the kinetic part of the Hamiltonian (including spin-orbit terms) for a general wave vector $\mathbf{k}$ is not diagonal. In order to be consistent with this convention, we translate the results of Sects. \ref{firstorder} and \ref{corrections}, formulated in the eigenstates of $\mathcal{H}_0\,$, to the spin-up/down basis. The unitary transformation connecting the two basis systems depends on the wave vector $\mathbf{k}$ and the type of spin-orbit interaction to be considered. To keep the discussion as general as possible, we take into account the two most frequently discussed mechanisms of spin-orbit coupling, namely the linearized Dresselhaus term and the Rashba spin-orbit interaction \cite{Winklerbook:2003,Schliemann:2003}. Accordingly, we have instead of $\mathcal{H}_0$ the Hamiltonian
\begin{equation}
\mathcal{H}_{\uparrow\, \downarrow}=\mathcal{H}_{kin} + \mathcal{H}_{R} + \mathcal{H}_{D}\; ,
\label{hamtrafo}
\end{equation}
with the kinetic energy $H_{kin} = \frac{\hbar^2}{2 m^*}\, k^2 \cdot \openone_{2\times2}$ of the free electron with effective mass $m^*\,$. The Rashba-Hamiltonian has the form $\mathcal{H}_R= \alpha\, (k_x\, \sigma_{y} - k_y\, \sigma_{x})$ with the Rashba coefficient $\alpha\,$, the Pauli spin matrices $\sigma_{x/y}$ and the components $k_{x/y}$ of the in-plane wave vector. The linearized Dresselhaus-Hamiltonian has a similar form, given by $\mathcal{H}_D=\beta\, (k_x\, \sigma_x - k_y \sigma_y)$ with the weighting parameter $\beta\,$. The appearance of the Pauli spin matrices in $\mathcal{H}_R$ and $\mathcal{H}_D$ indicates the use of a basis with spin orientation parallel (or antiparallel) to the $z$-axis (which usually is the growth direction of the QW). The unitary transformation we are looking for is obtained by diagonalizing $\mathcal{H}_{\uparrow\, \downarrow}$ (Eq. \eqref{hamtrafo}) to find the eigenvectors
\begin{equation}
|\pm \rangle_{\mathbf{k}} = \frac{1}{\sqrt{2}}\,\left(\begin{array}{c}
\pm A_{\mathbf{k}} \\ 1 
\end{array} \right)\; ,
\end{equation} 
with
\begin{equation}
A_{\mathbf{k}} = \frac{-\beta\, k_{+} + i \alpha k_{-}}{\sqrt{(\alpha^2 + \beta^2)(k_{x}^{2} + k_{y}^{2}) - 4\, \alpha\beta\, k_x\, k_y}}
\label{unitaryA}
\end{equation}
and the common abbreviation $k_{\pm}=k_x + i\, k_y\,$. 

Applying the transformation matrix, composed of these eigenvectors, to the density matrix $\bm{\varrho}^{(m_c\, \bar{m}_c)}(\mathbf{k})\,$, we get the spin-density matrix in the basis of the spin-up and spin-down states
\begin{widetext}
\begin{eqnarray}
\nonumber
\bm{\varrho}^{(\uparrow\, \downarrow)}(\mathbf{k}) &=&
\left(\begin{array}{c c}
\varrho_{\uparrow\, \uparrow}(\mathbf{k}) & \varrho_{\uparrow\, \downarrow}(\mathbf{k}) \\
\varrho_{\downarrow\,\uparrow}(\mathbf{k}) & \varrho_{\downarrow\, \downarrow}(\mathbf{k}) \end{array} \right)\\
&=& \frac{1}{2}\left(\begin{array}{c c} 
d_{+}(\mathbf{k}) + 2\Re\left\{ A^{*}_{\mathbf{k}}\, \varrho_{m_c\, -m_c}(\mathbf{k})\right\} & -d_{-}(\mathbf{k}) +2i\, \Im\left\{ A^{*}_{\mathbf{k}}\, \varrho_{m_c\, -m_c}(\mathbf{k})\right\}\\ \nonumber
-d_{-}(\mathbf{k}) - 2i\, \Im\left\{A^{*}_{\mathbf{k}} \,\varrho_{m_c\, -m_c}(\mathbf{k}) \right\} &
d_{+}(\mathbf{k}) - 2\, \Re\left\{A^{*}_{\mathbf{k}}\, \varrho_{m_c\,-m_c}(\mathbf{k}) \right\} 
 \end{array} \right)\; , \\
\label{unitarymatrix}
\end{eqnarray}
\end{widetext}
with $d_{\pm}(\mathbf{k})=\varrho_{m_c\, m_c}(\mathbf{k}) \pm \varrho_{-m_c\, -m_c}(\mathbf{k})\,$. For a particular choice of the spin-orbit interaction (Rashba or Dresselhaus) the corresponding unitary transformation can be derived on the basis of this result.

Spin-dynamics experiments, such as time-resolved photoluminescence  or Faraday rotation (for an overview of recent experiments using these techniques see Ref. \onlinecite{Awschalombook:2002}) or photogalvanic effect \cite{Ganichev:2002} do not aim at the dynamics of the density matrix of an individual $\mathbf{k}$ but at quantities such as the \emph{spin polarization}
\begin{equation}
S = \sum_{\mathbf{k}}\left(\varrho_{\uparrow\, \uparrow}(\mathbf{k}) - \varrho_{\downarrow\, \downarrow}(\mathbf{k}) \right)
\end{equation}
and the \emph{spin coherence} \cite{Weng:2003}
\begin{equation}
C = \sum_{\mathbf{k}}|\varrho_{\uparrow\, \downarrow}(\mathbf{k})|\, ,
\end{equation}
defined for the whole population of the two-level system. Their decay is characterized by the spin relaxation time $\tau_{SR}$ and the spin decoherence time $\tau_{SD}\,$. With the results of Sects. \ref{firstorder} and \ref{corrections} we are now in the state to formulate the relation between these quantities and the relaxation times $T_{1,\, \mathbf{k}}$ and $T_{2,\, \mathbf{k}}$ by applying the unitary transformation to express
\begin{eqnarray}
S &=& \sum_{\mathbf{k}} 4 \Re\left\{A^{*}_{\mathbf{k}}\, \varrho_{m_c\, -m_c}(\mathbf{k}) \right\} \label{Sdef}\\
C &=& \sum_{\mathbf{k}} \sqrt{d_{-}^{2} + 4 \Im\left\{A^{*}_{\mathbf{k}}\, \varrho_{m_c\, -m_c}(\mathbf{k}) \right\}}\label{Cdef}\; .
\end{eqnarray}
The time-derivatives of $S$ and $C$ depend on those of the original pseudospin-density matrix $\varrho^{(m_c\, \bar{m}_c)}(\mathbf{k})\,$. Thus, the decay times of $S$ and $C\,$, i.e. the spin relaxation and the spin decoherence are determined by  $T_{1,\, \mathbf{k}}$ and $T_{2,\, \mathbf{k}}$ derived in the previous sections, yet in a complicated relation. A microscopic calculation of $\tau_{SR}$ and $\tau_{SD}$ has to make use of this relation. Nevertheless, it is possible to state, that the decay of the spin polarization $S$ is determined only by $\varrho_{m_c\, -m_c}(\mathbf{k})\,$, i.e. by the transverse pseudospin relaxation time $T_{2,\, \mathbf{k}}\,$, while the decay of the spin coherence $C$ depends on both the longitudinal \emph{and} transverse pseudospin relaxation times $T_{1,\, \mathbf{k}}$ and $T_{2,\, \mathbf{k}}\,$ \footnote{The results in Ref. \onlinecite{Weng:2003} are restricted to the spin coherence $C$ but do not touch the relation with $T_{1,\, \mathbf{k}}$ and $T_{2,\, \mathbf{k}}\,$, to which we focus here}. The existence of a complicated relation between $T_{1,\, \mathbf{k}}\,$, $T_{2,\, \mathbf{k}}$ for a simple spin-split two-level system and the spin polarization and spin coherence decay times of a whole carrier population has been mentioned before in the literature (see  chapter 4 of Ref. \onlinecite{Awschalombook:2002}) but without making it explicit.  

%%%%%%%%%%%%%%%%%%%%%%%%%%%%%%%%%%%%%%%%%%%%%%%%%%%%%%%%%%%%%%%%%%%%%%%%%%%%%%%%%%%%%%%%%%%%%%%%%%%%%%%%%%%

\section{\label{summary}Conclusions}

In this paper we have presented a microscopic formulation of spin dynamics in semiconductor heterostructures. It is based on the density matrix approach and its particular form, the optical Bloch equations. Starting from the 6-level system of conduction and valence band states driven by optical excitation and including carrier-phonon interaction we derive explicitly the EOM for the $2 \times 2$ density matrix of the electron subsystem whose energy levels are spin-split due to spin-orbit coupling. We employ a truncation scheme to include electron-phonon interaction in second order. In this limit we derive microscopic expressions for the longitudinal and transverse (pseudo-) spin relaxation times for the individual spin-split two-level system at a fixed $\mathbf{k}\,$. Finally a connection between these results  and spin relaxation times characterizing the dynamics of a whole population and accessible by experiments is established. It takes into account the different sets of eigenstates used in our microscopic derivation (which diagonalizes the spin-orbit coupling) and in the interpretation of the measurable times (with a fixed axis for spin quantization and nondiagonal spin-orbit coupling). Thus we provide at the same time a microscopic formulation of spin dynamics and its relation to experiments. 

We would like to emphasize that the concept presented here can be extended to include also carrier-carrier interaction thus arriving at an extension of the coherent SBE of Ref. \onlinecite{Roessler:2002}. For preliminary results we refer to Ref. \onlinecite{Lechner:2004b}. Further steps will be numerical evaluations of the microscopic expressions for realistic quantum structures and the explicit treatment of the spin dynamics for the hole system.
 
%%%%%%%%%%%%%%%%%%%%%%%%%%%%%%%%%%%%%%%%%%%%%%%%%%%%%%%%%%%%%%%%%%%%%%%%%%%%%%%%%%%%%%%%%%%%%%%%%%%%%%%%%%%

\section{\label{acknowledgement}Acknowledgment}

We thankfully acknowledge financial support from the DFG via Forschergruppe 320/2-1 ``Ferromagnet-Halbleiter-Nanostrukturen''.

%end of main text
%
%%%%%%%%%%%%%%%%%%%%%%%%%%%%%%%%%%%%%%%%%%%%%%%%%%%%%%%%%%%%%%%%%%%%%%%%%%%%%%%%%%%%%%%%%%%%%%%%%%%%%%%%%%%
%begin of inclusion of BibTeX-File

\bibliography{promotion}

%end of inclusion of BibTeX-File
%
%%%%%%%%%%%%%%%%%%%%%%%%%%%%%%%%%%%%%%%%%%%%%%%%%%%%%%%%%%%%%%%%%%%%%%%%%%%%%%%%%%%%%%%%%%%%%%%%%%%%%%%%%%%

\end{document}